\documentclass[11pt]{article}
\usepackage{moriond,epsfig}

\def\be{\begin{equation}}
\def\ee{\end{equation}}
\def\bea{\begin{eqnarray}}
\def\eea{\end{eqnarray}}

\begin{document}
\vspace*{4cm}
\title{EXOTIC SEARCHES AT ATLAS}

\author{ D.M. GINGRICH\\(on behalf of the ATLAS Collaboration)}

\address{Centre for Particle Physics, Department of Physics, University
of Alberta, \\ Edmonton, AB T6G 2G7 Canada \&\\  
TRIUMF, Vancouver, BC V6T 2A3 Canada}\

\maketitle\abstracts{
We present the first results of searches for new physics with the ATLAS
detector using the 2010 Large Hadron Collider proton-proton collision
data at a centre of mass energy of 7~TeV. 
After a few months of operation, these searches already go beyond the
reach of previous experiments, and start to explore new territories.}

\section{Introduction\label{sec1}}

This paper presents five searches for new physics in proton-proton
collisions using the ATLAS detector at the Large Hadron Collider.
The data were collected in 2010 at a centre of mass energy of 7~TeV.
The first two searches use 3.1~pb$^{-1}$ of early data, while the later
three searches use the full 2010 data set with a typical luminosity of
36~pb$^{-1}$.   

\section{Long-Lived Highly Ionising Particles}

The ATLAS collaboration has performed a search for massive long-lived
highly ionising particles (HIP).~\cite{HIP} 
Some examples that may give rise to highly ionising particle signatures
are Q-balls, black hole remnants, magnetic monopoles, and dyons.
We have performed a model independent search.
Due to their large mass, HIPs are also characterised by their
non-relativistic speeds, as well as, high electric charge.
We expect large amounts of energy loss through ionisation for these
states. 
In ATLAS, HIPs would leave tracks in the inner tracking detector,
matched to narrow energy loss in the electromagnetic calorimeter.

ATLAS is not able to search for HIP of all charges, masses,  and
lifetimes.
The accessible parameter space was determined as follows.
A lower charge bound of $|q| \ge 6e$ was determined by the
$E_\mathrm{T} > 10$~GeV trigger threshold. 
The upper charge bound of $|q| \le 17e$ was determined by delta
electrons and electron recombination.
An upper bound on the mass of 1~TeV was determined by trigger timing
constraints.  
A lifetime greater than 100~ns was required to maintain narrow energy
deposits.   
A data sample with a luminosity of 3.1~pb$^{-1}$ was used.

\begin{figure}[htb]
\begin{center}
\includegraphics[width=9cm]{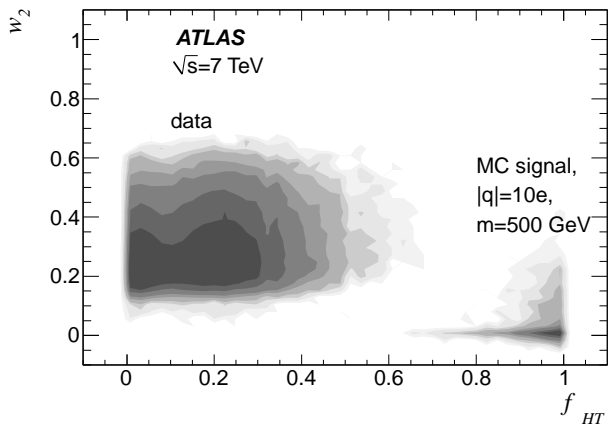}
\caption{Contours of $w_2$ versus $f_{HT}$ distributions showing the
density of entries on a log scale. 
Data and a signal Monte Carlo simulated sample are shown.
\label{fig01}}
\end{center}
\end{figure}

HIPs were discriminated by the proportion of high-ionisation hits and the 
lateral extent of the energy deposition.
Specifically, the fraction $f_{HT}$ of transition radiation tracker
(TRT) hits on the track which pass a high ionisation threshold was used.
In addition, a requirement on the fraction of energy outside the three
most energetic cells associated to a selected electromagnetic (EM)
energy cluster, in the second EM calorimeter layer, $w_2$, was made.  
Figure~\ref{fig01} shows that the data matches Standard Model (SM)
expectations, and no HIPs were observed.
The estimated background in the signal region was $0.019\pm 0.005$ events.
Limits for particles produced in the acceptance kinematic region and by
Drell-Yan production are shown in Table~\ref{tab1}.
A Bayesian statistical approach with a uniform prior for the signal was
used. 
HIP masses above 800~GeV are probed for the first time at particle
colliders.
 
\begin{table}[htb]
\caption{Inclusive and pair production cross section upper limits
(95\% C.L.) for long-lived massive particles with high electric
charges $|q|$, produced in the search acceptance and assuming a
Drell-Yan production mechanism. 
\label{tab1}}
\vspace{0.4cm}
\begin{center}
\begin{tabular}{|c|ccc|ccc|}\hline
Mass & \multicolumn{3}{c|}{Inclusive Search} &
\multicolumn{3}{c|}{Drell-Yan Mechanism}\\\cline{2-7} 
& $|q| = 6e$ & $|q| = 10e$ & $|q| = 17e$ & $|q| = 6e$ & $|q| =
10e$ & $|q| = 17e$\\\hline
 200~GeV & 1.4~pb & 1.2~pb & 2.1~pb & 11.5~pb & 5.9~pb & 9.1~pb\\
 500~GeV & 1.2~pb & 1.2~pb & 1.6~pb & \ 7.2~pb & 4.3~pb & 5.3~pb\\
1000~GeV & 2.2~pb & 1.2~pb & 1.5~pb & \ 9.3~pb & 3.4~pb & 4.3~pb\\
\hline
\end{tabular}
\end{center}
\end{table}


\section{Diphoton with Large Missing Energy}

ATLAS has performed a search for events with diphotons ($\gamma\gamma$)
and large missing transverse energy
$E_\mathrm{T}^\mathrm{miss}$.~\cite{UED}  
This signature has been interpreted in the context of Universal Extra
Dimensions (UED).  
We considered a single TeV$^{-1}$ sized UED with a compactification
radius $R$.  
In this model, the lightest Kaluza-Klein (KK) particle (LKP) is the KK
photon $\gamma^*$. 
The KK particles are produced as pairs of KK quarks and/or KK gluons in
the strong interaction.
These KK particles then decay down, via KK states, to the LKP.
The LKP decays by $\gamma^* \to \gamma + G$.
We interpreted the results of the search using a model in which $\Lambda
R = 20$, where $\Lambda$ is the UV cutoff and $R$ is a free 
parameter 

\begin{figure}[htb]
\begin{minipage}[t]{7.8cm}
\includegraphics[width=7.8cm]{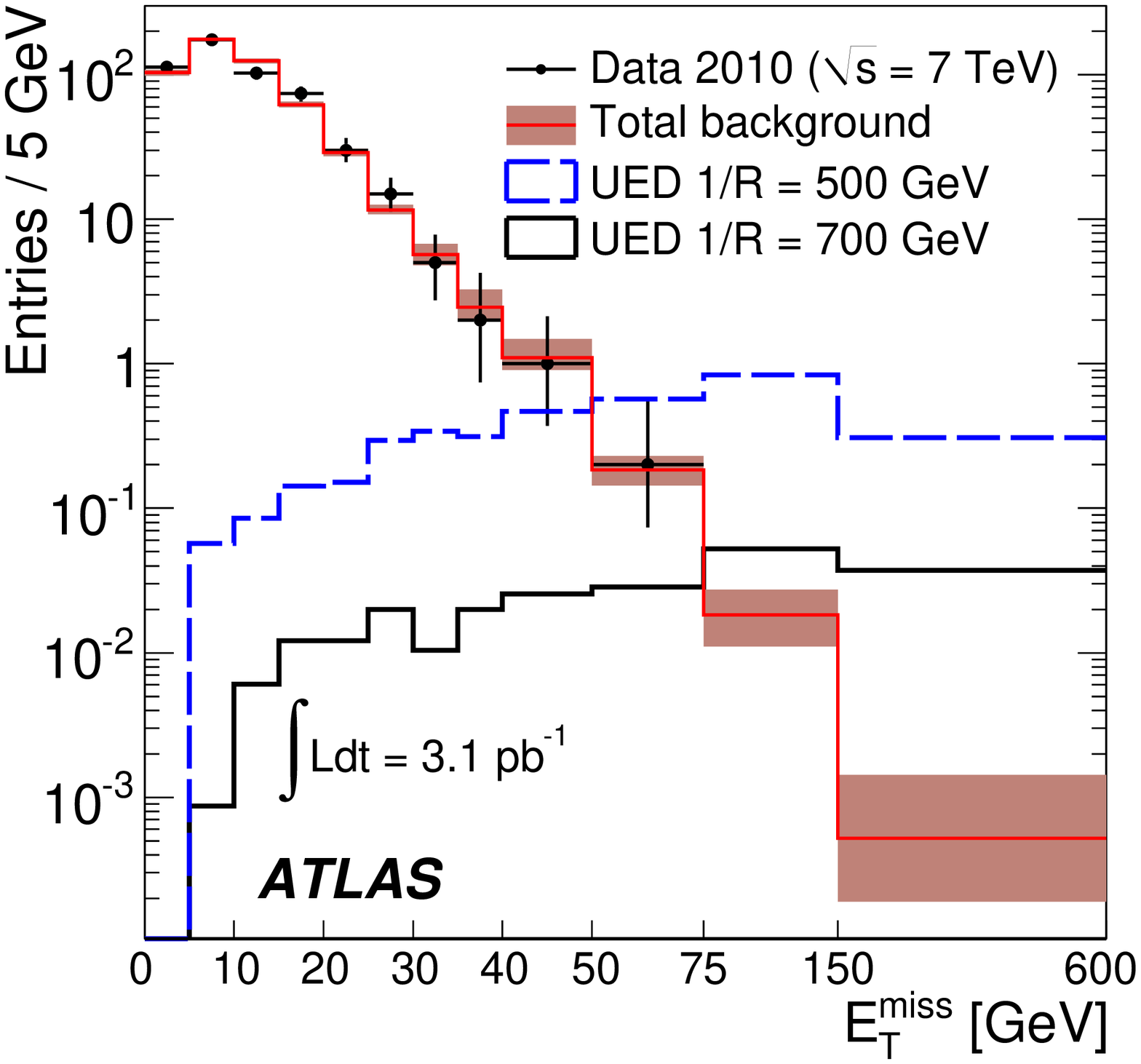}
\caption{$E_\mathrm{T}^\mathrm{miss}$ spectra for $\gamma\gamma$ events,
compared to the total SM background as estimated from data.  
Also shown are two hypothetical UED signals.
The vertical error bars and shaded bands show the statistical errors.
\label{fig02}}
\end{minipage}
\hfill
\begin{minipage}[t]{7.8cm}
\includegraphics[width=7.8cm]{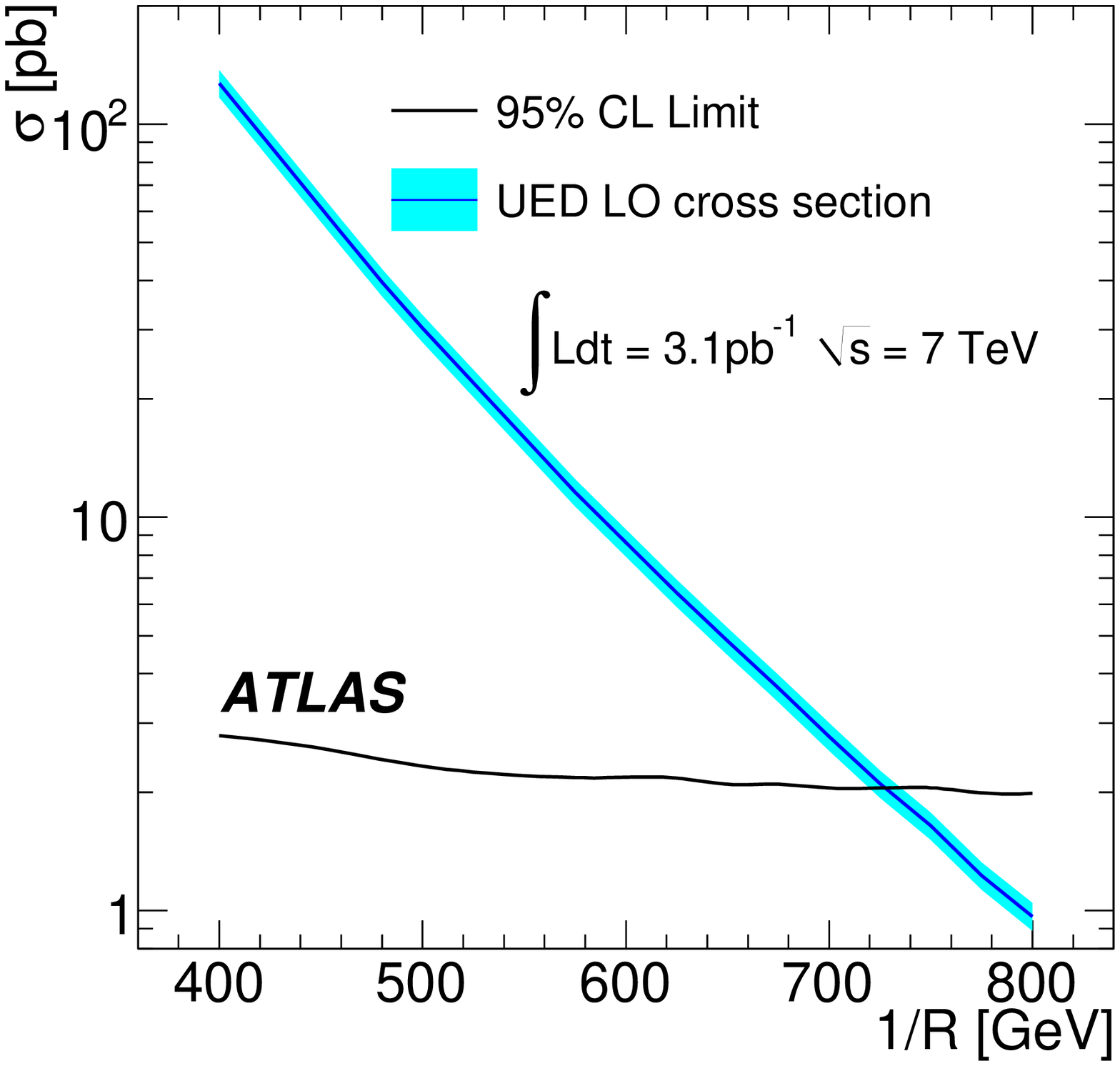}
\caption{95\% C.L. upper limits on the UED production cross section,
and the leading order (LO) theory cross section prediction, as a
function of $1/R$. 
The shaded band shows the PDF uncertainty.
\label{fig03}}
\end{minipage}
\end{figure}

Figure~\ref{fig02} shows the $E_\mathrm{T}^\mathrm{miss}$ spectrum of
events with diphotons.
Events were required to have two photons each with $E_\mathrm{T} >
25$~GeV, and an event $E_\mathrm{T}^\mathrm{miss} > 75$~GeV. 
Zero signal events were observed and the estimated background was
$0.32\pm 0.16 ^{+0.37}_{-0.10}$ events. 
Figure~\ref{fig03} shows upper limits on the cross section.
The upper limits were calculated using a Bayesian approach with a flat
prior for the signal cross section.
It was verified that the result is not very sensitive to the detailed
form of the assumed prior.
In context of the previously specified model, values of $1/R < 728$~GeV
are excluded.
This is the most sensitive limit on this model to date.

\section{Search for New Physics in Dijets}

\begin{figure}[b!]
\begin{center}
\includegraphics[width=7.8cm]{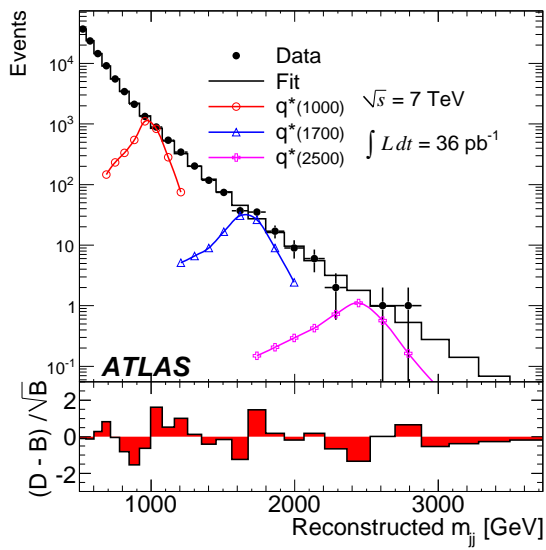}
\caption{Observed (D) dijet mass distribution (solid dots) fitted using
a binned QCD background (B) parameterisation (histogram). 
Predicted q$^*$ signals normalised to 36~pb$^{-1}$ for masses of
1.0, 1.7, and 2.5~TeV are overlaid. 
The bin-by-bin significance of the data-background difference is shown
in the lower panel.
\label{fig04}}
\end{center}
\end{figure}

ATLAS has performed a study of dijet events using both the invariant
mass of the two jets and angular distributions of energetic jets up
to 3.5~TeV.~\cite{dijet} 
For the invariant mass studies, we required $p_\mathrm{T}^\mathrm{j_1} >
150$~GeV and $p_\mathrm{T}^\mathrm{j_2} > 30$~GeV, as well as,
$|\Delta\eta_\mathrm{jj}| > 1.3$.
Figure~\ref{fig04} shows that the invariant mass distribution is smooth
as expected for QCD jet production and agrees with the SM background
parameterisation. 

\begin{figure}[htb]
\begin{minipage}[t]{7.8cm}
\includegraphics[width=7.8cm]{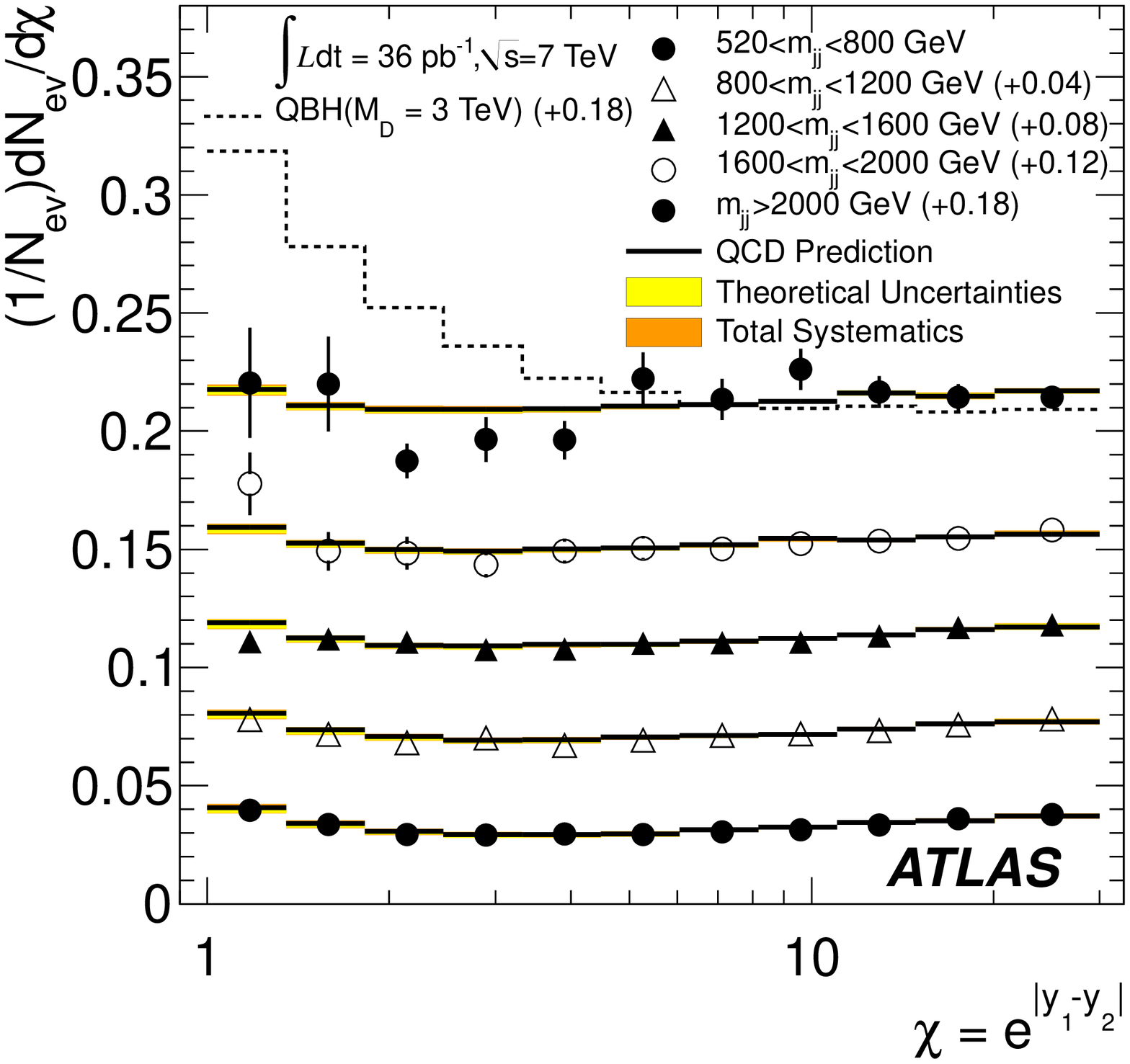}
\caption{$\chi$ distributions for different $m_\mathrm{jj}$ bins.
Shown are the QCD predictions with systematic uncertainties, and data
points with statistical uncertainties. 
The dashed line is the prediction of a quantum black hole (QBH) signal
in the highest mass bin. 
The distributions and QCD predictions have been offset.
\label{fig05}}
\end{minipage}
\hfill
\begin{minipage}[t]{7.8cm}
\includegraphics[width=7.8cm]{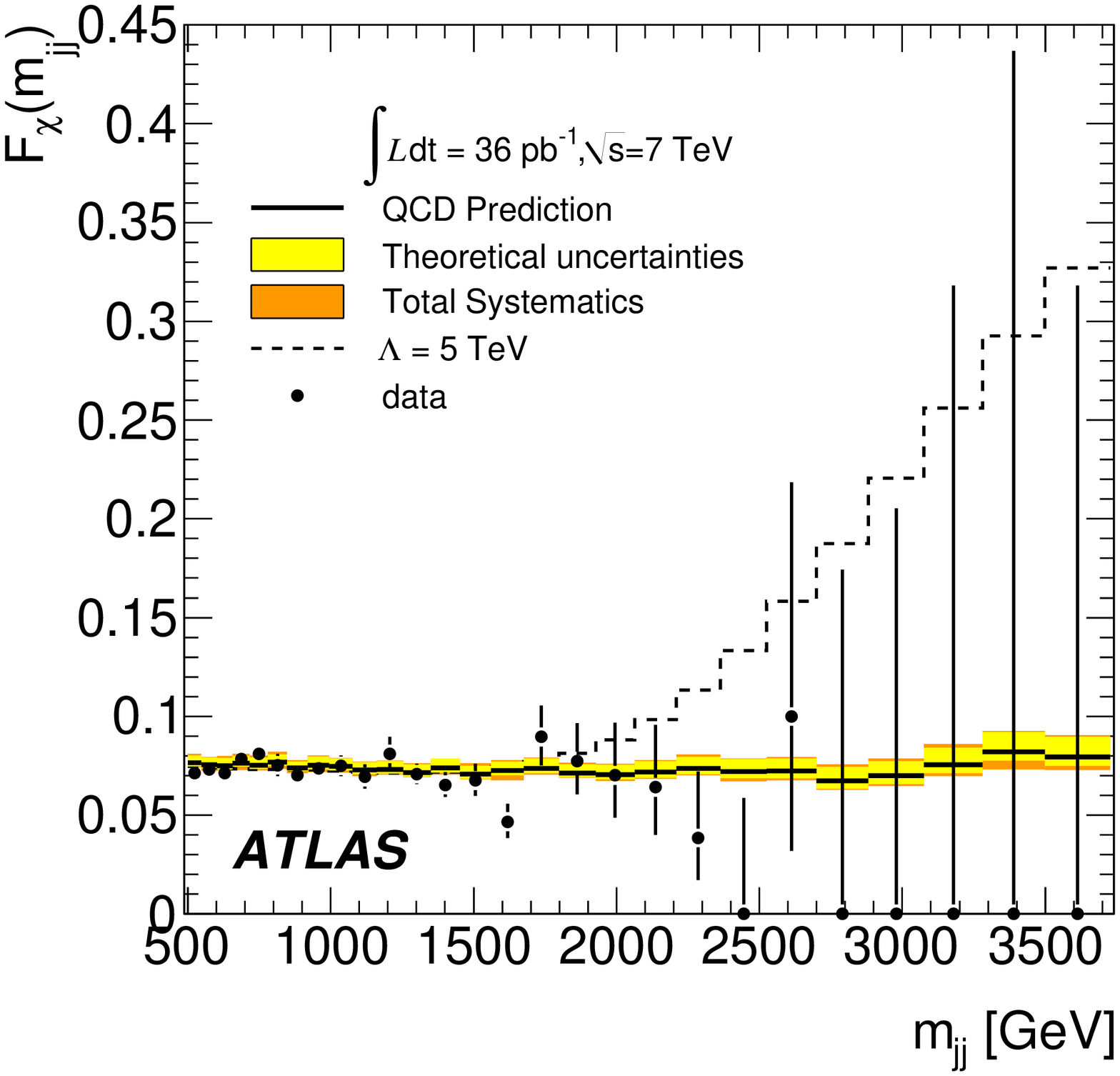}
\caption{$F_\chi(m_\mathrm{jj})$ function versus $m_\mathrm{jj}$.
We show the QCD prediction with systematic uncertainties (band) and data
(solid dots) with statistical uncertainties.
The expected signal for QCD plus a quark contact interaction with
$\Lambda = 0.5$~TeV is also shown.
\label{fig06}}
\end{minipage}
\end{figure}

For the angular distributions, we required $p_\mathrm{T}^\mathrm{j_1} >
60$~GeV and $p_\mathrm{T}^\mathrm{j_2} > 30$~GeV. 
The rapidities of the two leading jets per event are required to satisfy
$y_B = 0.5 (y_1 + y_2) < 1.10$ and $y^* = 0.5(y_1 - y_2) < 1.70$. 
Figure~\ref{fig05} shows the $\chi$ distributions, where $\chi =
\exp(|y_1-y_2|) = \exp(2|y^*|)$. 
Data are consistent with QCD.
We also examined the dijet centrality, where $F_\chi(m_\mathrm{jj}) = 
N_\mathrm{events}(|y^*| < 0.6)/N_\mathrm{events}(|y^*| < 1.7)$. 
This distribution is shown in Fig.~\ref{fig06}.

We now interprete the results using several models, showing both the
dijet mass and angular distribution results.
For the resonance results, we set Bayesian credibility levels by
defining a posterior probability density from the likelihood function
for the observed mass spectrum, obtained by a fit to the background functional
form and a signal shape derived from MC calculation.
For the angular distribution results, likelihood ratios for comparing
the different hypotheses and parameter estimators were used.
Confidence level limits are set using the frequentist CLs+b approach. 

\begin{figure}[p]
\begin{minipage}[t]{7.8cm}
\includegraphics[width=7.8cm]{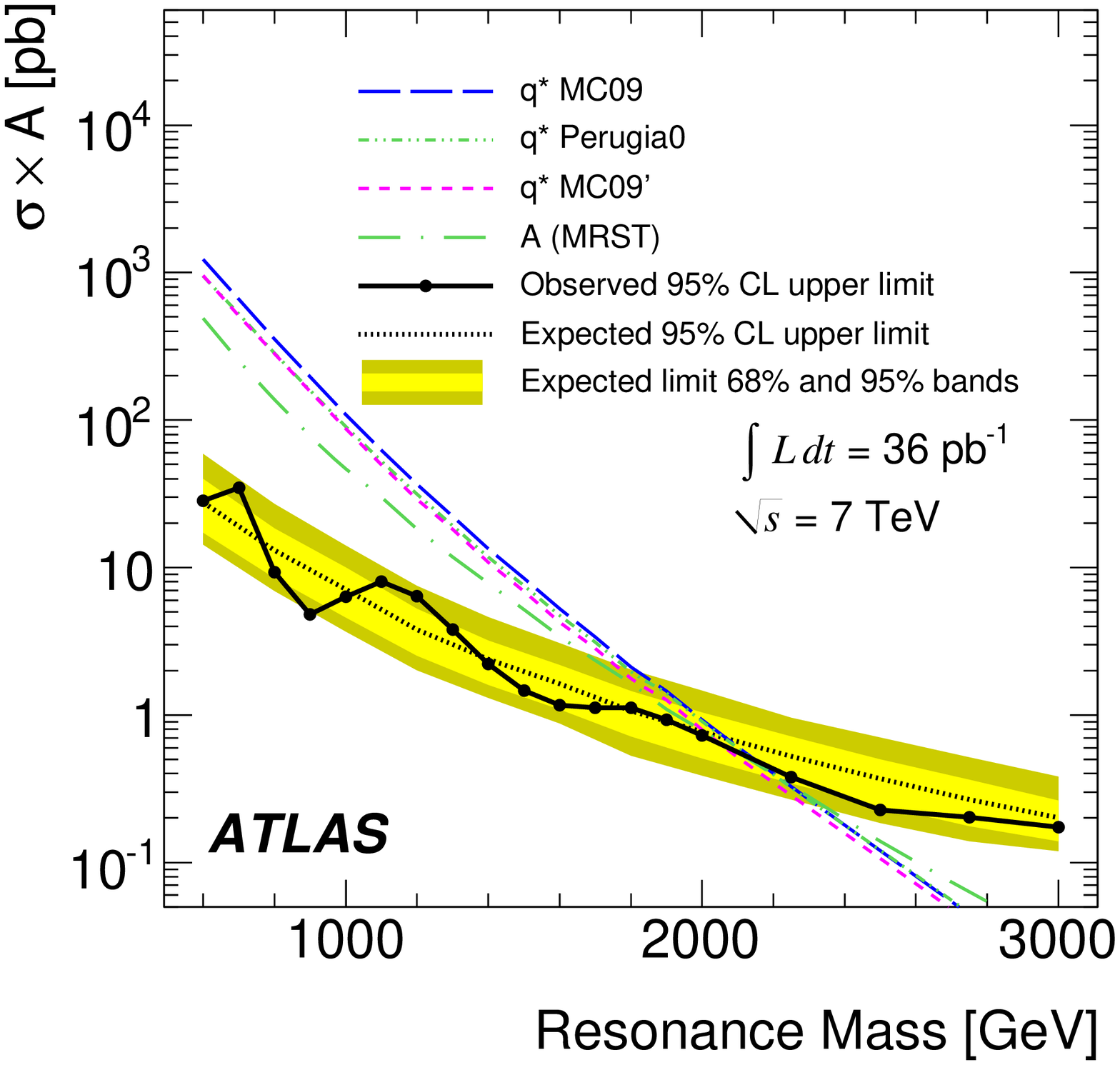}
\caption[]{95\% C.L. upper limits on cross section times
acceptance for a resonance decaying to dijets taking into account both
statistical and systematic uncertainties (points and solid line)
compared to an axigluon model and to a q$^*$ model with three
alternative MC tunes.
\label{fig07}}
\end{minipage}
\hfill
\begin{minipage}[t]{7.8cm}
\includegraphics[width=7.8cm]{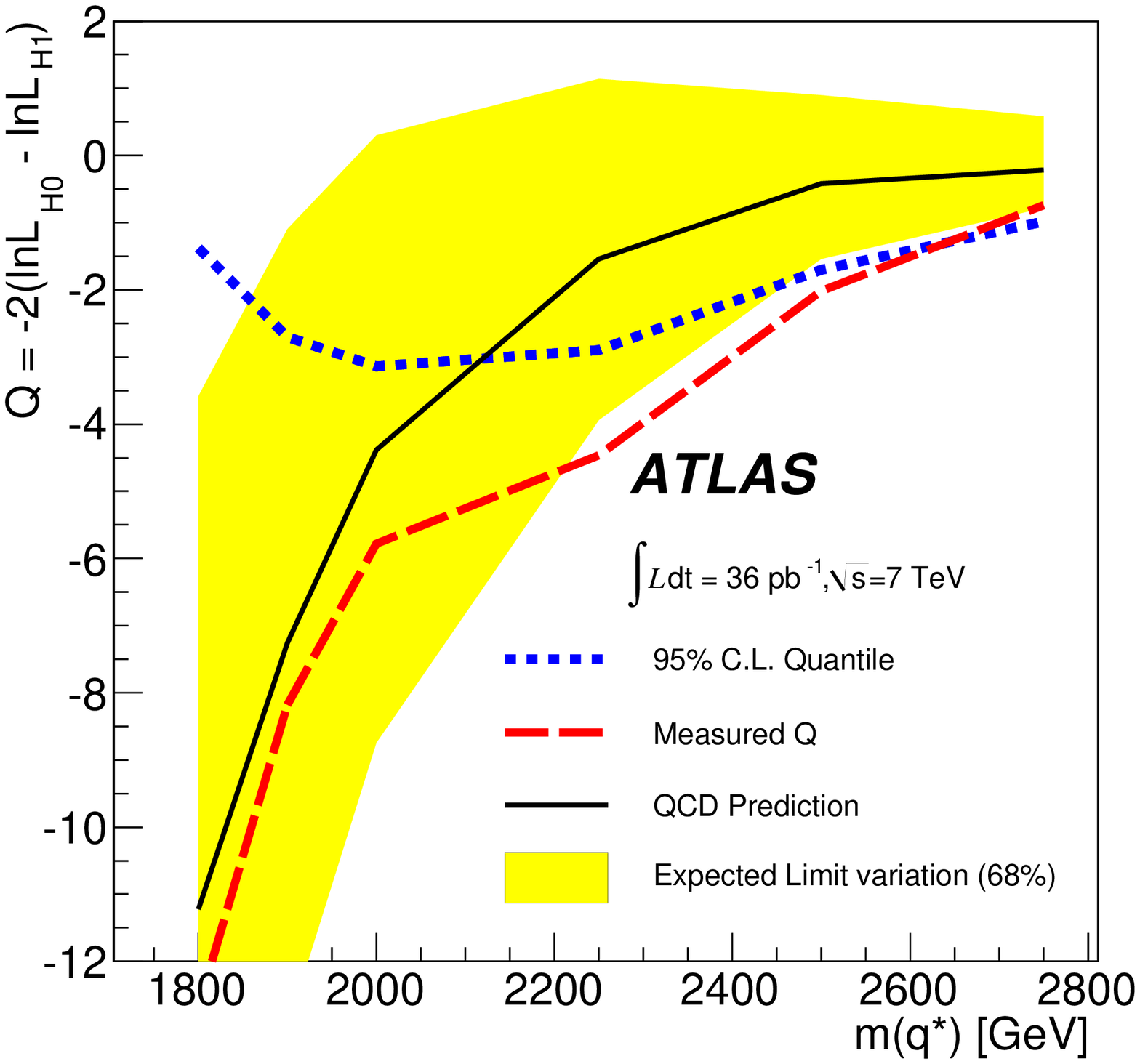}
\caption[]{95\% C.L. limits on the excited quark model using the
logarithm of the likelihood ratios obtained from the
$F_\chi(m_\mathrm{jj})$ distributions. 
The expected 68\% interval for the expected limits are shown by the
band.
\label{fig08}}
\end{minipage}
\end{figure}

Excited quarks can be produced in $\mathrm{qg \to q^*}$ and decay by
$\mathrm{q^*\to q g}$, $\mathrm{q W/Z}/\gamma$. 
Figure~\ref{fig07} shows the results of the resonance search.
Excited quarks are excluded in the mass range $0.60 < m < 2.15$~TeV,
while axigluons are excluded in the range $0.60 < m < 2.10$~TeV.
Shown in Fig.~\ref{fig08} is the $Q$ distribution, where
$Q = -2[\ln(F_\chi(m_\mathrm{jj})|H0) - \ln(F_\chi(m_\mathrm{jj})|H1)
]$, and $H0$ is 
the null hypothesis (QCD only) and $H1$ is the hypothesis for new physics.
From this analysis, excited quarks in the mass range $0.60 < m <
2.64$~TeV are excluded.  

\begin{figure}[p]
\begin{minipage}[t]{7.8cm}
\includegraphics[width=7.8cm]{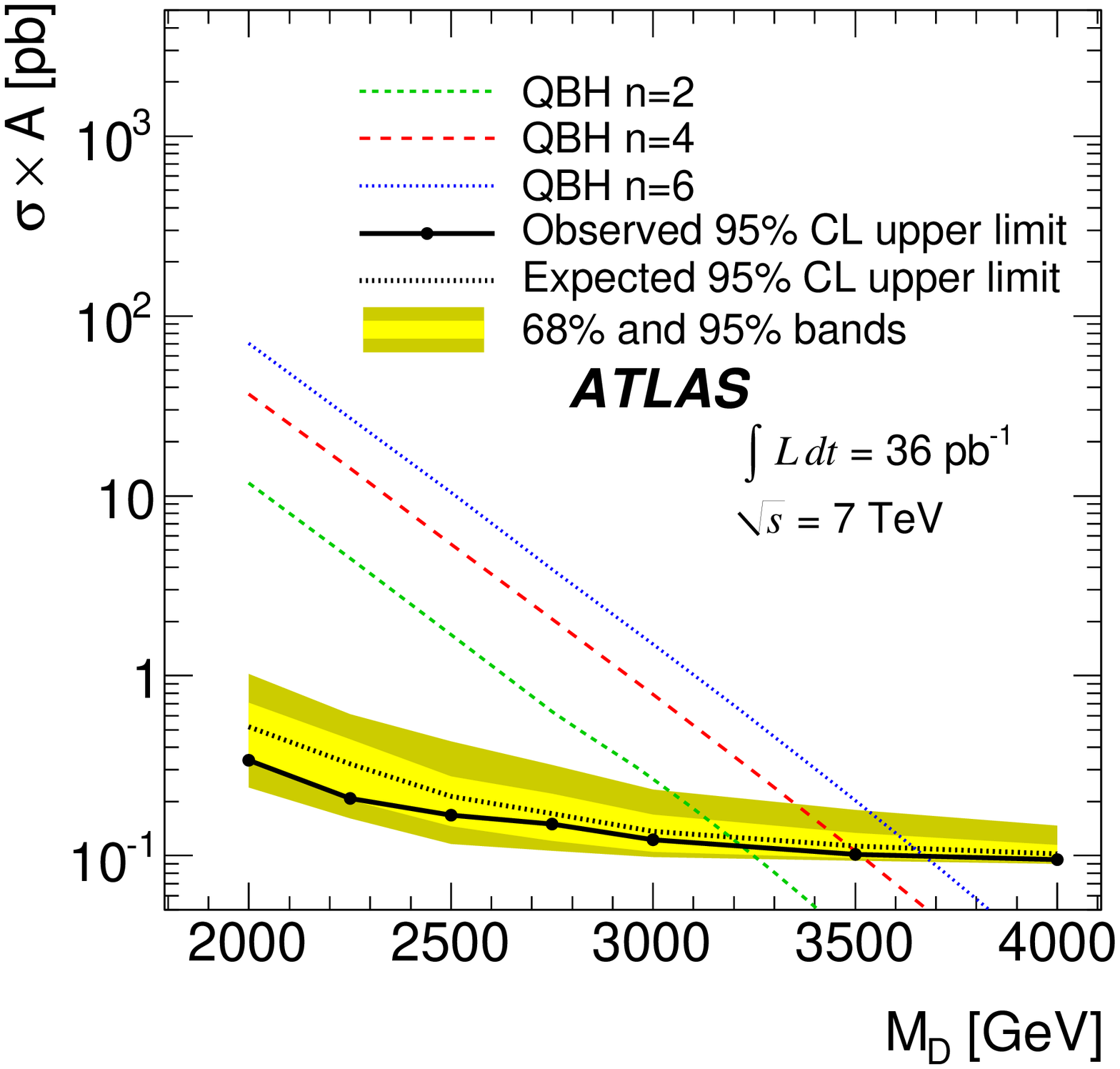}
\caption[]{95\% C.L. limits on the cross section times acceptance
versus the Planck scale for three quantum black hole (QBH) models,
taking into account both statistical and systematic uncertainties. 
\label{fig09}}
\end{minipage}
\hfill
\begin{minipage}[t]{7.8cm}
\includegraphics[width=7.8cm]{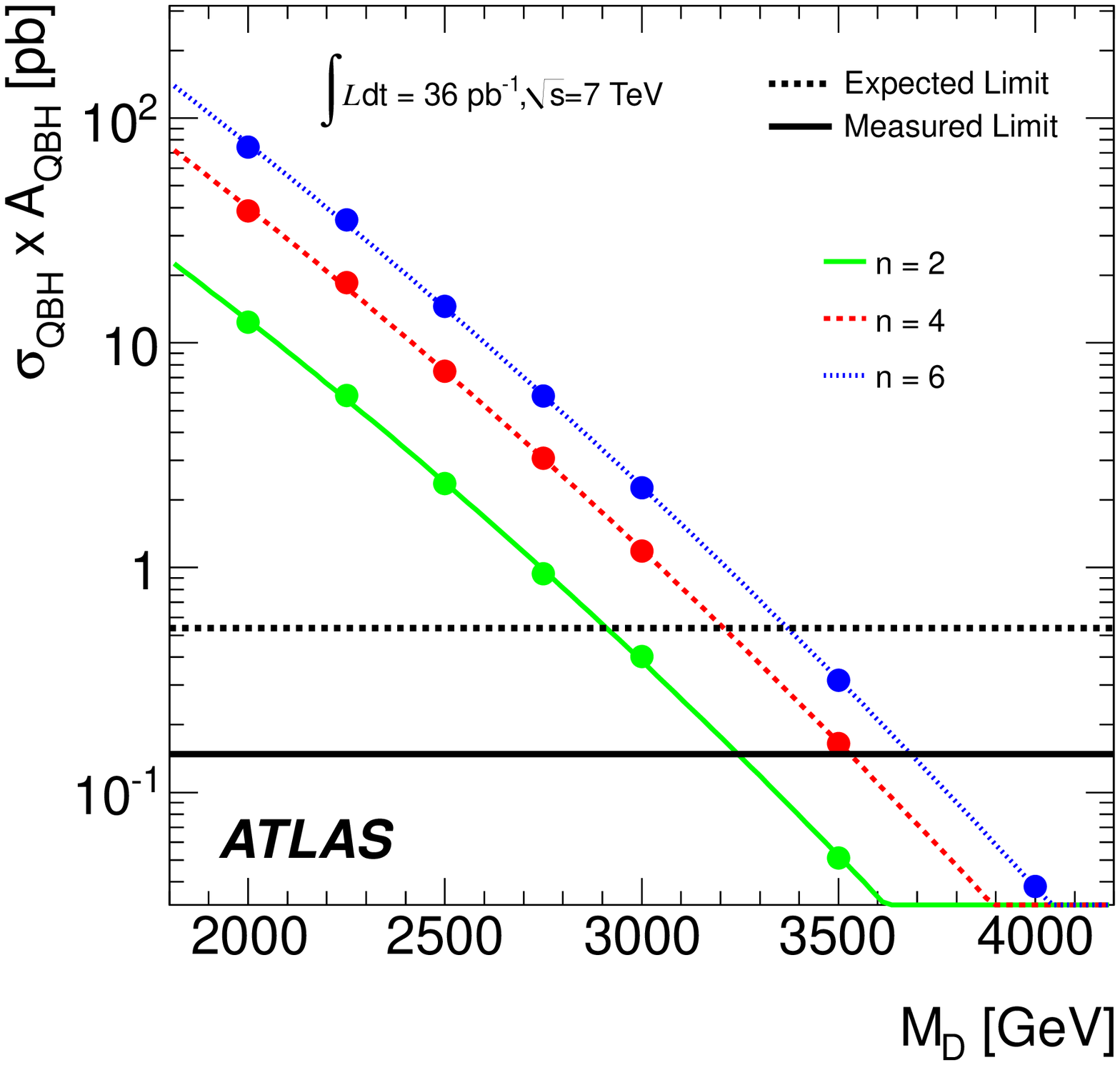}
\caption[]{Cross section times acceptance for quantum black holes (QBH)
as a function of $M_D$.
The measured and expected limits are shown as the solid and dashed lines.
\label{fig10}}
\end{minipage}
\end{figure}

We searched for quantum black holes (QBH) decaying to dijets, where
$M_D$ is the higher-dimensional Planck scale and $n$ is the number of
extra dimensions. 
These states would be expected to produce a large mass threshold effect
with long tails to higher masses. 
The results of the resonance search are show in Fig.~\ref{fig09}.
Planck scales in the range $0.75 < M_D < 3.67$~TeV are exclude.
The results of the angular distributions analysis are shown in
Fig.~\ref{fig10}.
From the $dN/d\chi$ distribution, Planck scales less than 3.69~TeV are
excluded, while from the $F_\chi(m_\mathrm{jj})$ distribution Planck
scales of less than 3.78~TeV are excluded.

\begin{figure}[htb]
\begin{center}
\includegraphics[width=7.8cm]{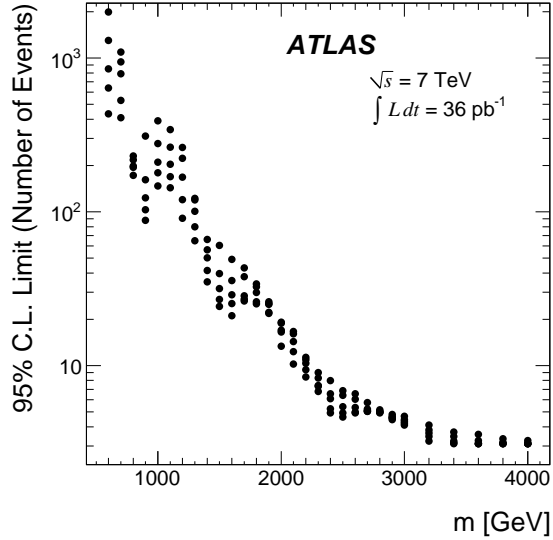}
\caption[]{95\% C.L. upper limits for the number of observed events
for Gaussians of width $\sigma/m$ of 0.03, 0.05, 0.07, 0.10, 0.15, at
each of various masses $m$. 
\label{fig11}}
\end{center}
\end{figure}

Finally, limits are given for a generic signal with a Gaussian profile.
Signal templates in the range $0.6 < m < 4.0$ with $3\% < \sigma < 15\%$
(5 different $\sigma$ values) were generated.
The results are shown in Fig.~\ref{fig11}.
These results can be used for different models by employing the following
prescription:
1) Check the validity of the Gaussian signal approximation, and
determine the peak and width of the signal;
2) Determine the model acceptance;
3) Calculate the event yield for the model cross section and luminosity of
36~pb$^{-1}$; 
4) Compare this event yield with the limits in Fig.~\ref{fig11}.

\section{Lepton plus Missing Transverse Energy}

\begin{figure}[htb]
\begin{center}
\includegraphics[width=7.8cm]{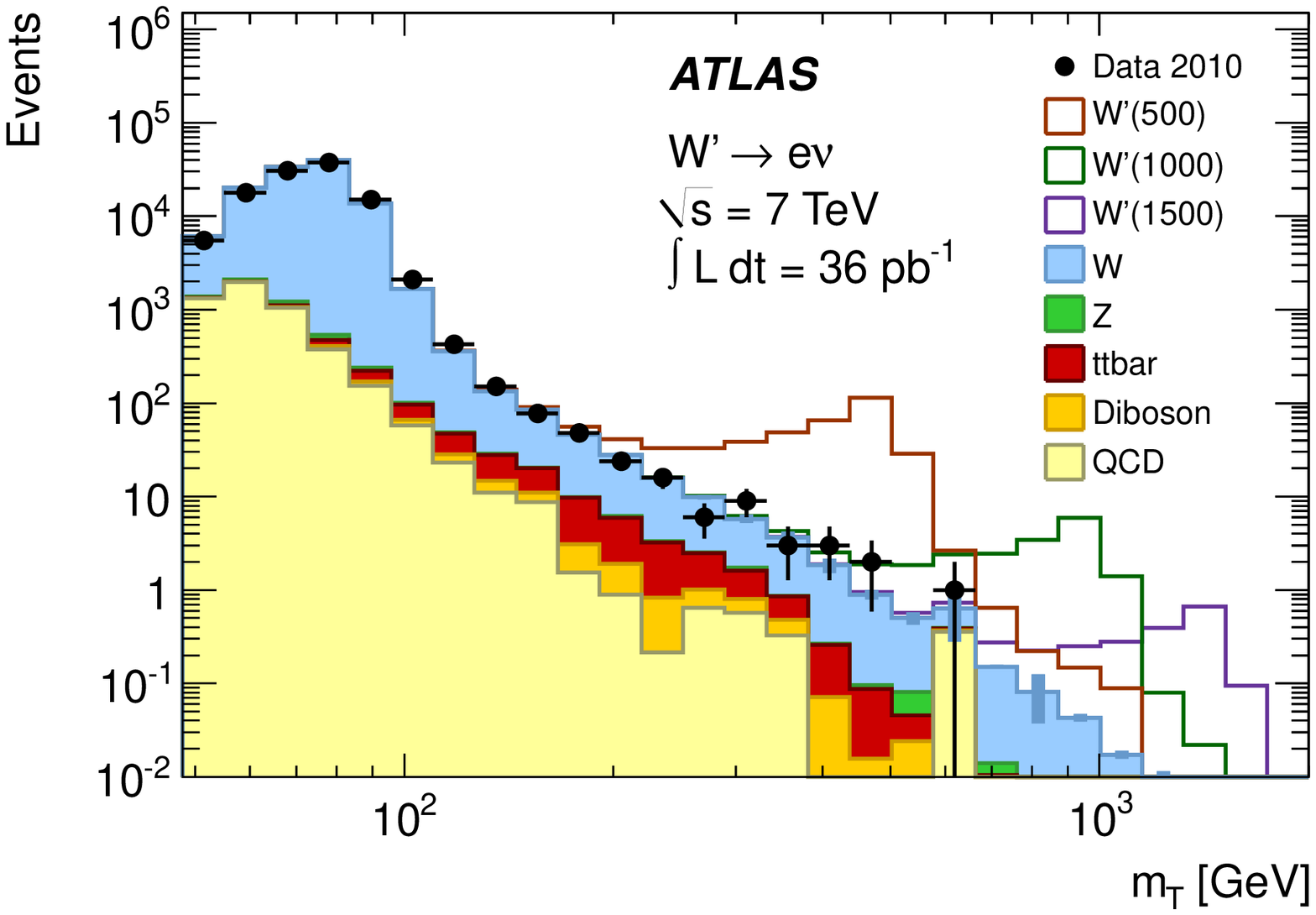}
\includegraphics[width=7.8cm]{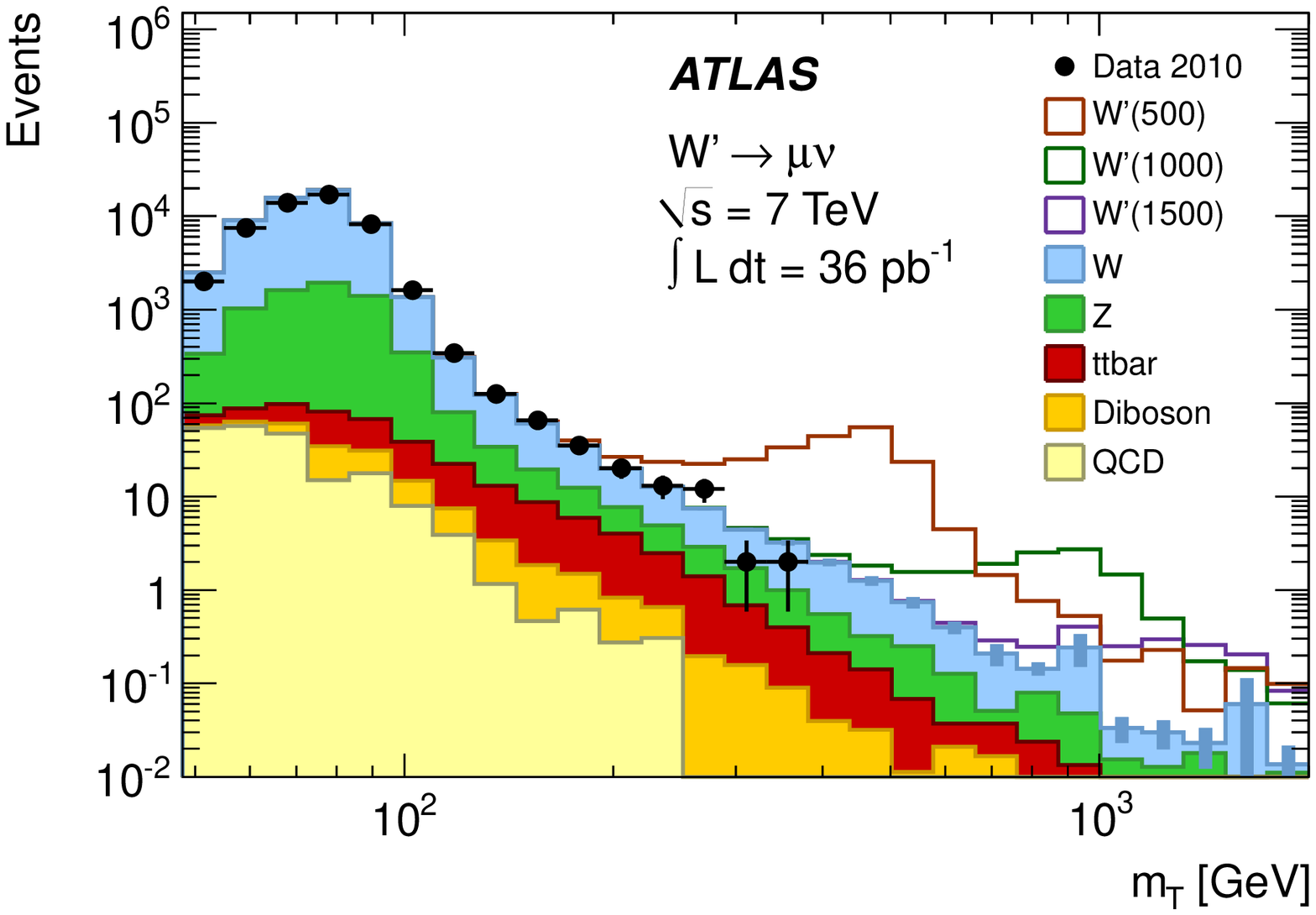}
\caption[]{$m_\mathrm{T}$ spectra for electron (left) and muon
(right) channels after final event selection.
The points represent data and the filled histograms show the
stacked backgrounds.
Open histograms are W$^\prime$ signals added to the background with
masses indicated in parentheses in the legend.
The signal and other background samples are normalised using the
integrated luminosity of the data and the NNLO (near-NNLO for
$\mathrm{t\bar{t}}$) cross sections.
\label{fig12}}
\end{center}
\end{figure}

ATLAS has performed a search for high-mass states decaying to an
electron or muon with missing energy:
W$^\prime/\mathrm{W}^* \to (\mathrm{e}/\mu) \nu$.~\cite{Wprime} 
The W$^\prime$ is a sequential SM boson with the same SM couplings as
the W-boson. 
The W$^*$ is a boson with anomalous magnetic moment type couplings.
The search was performed in the transverse mass defined as $m_\mathrm{T}
= \sqrt{2p_T E_\mathrm{T}^\mathrm{miss} (1-\cos\phi_{\ell\nu})}$. 
Events with electrons were chosen by requiring the electron to have
$E_\mathrm{T} > 25$~GeV, and the event to have $E_\mathrm{T}^\mathrm{miss} >
25$~GeV and  $E_\mathrm{T}^\mathrm{miss}/E_\mathrm{T} > 0.6$.  
Events with muons were chosen by requiring muons, in the barrel only, to
have $p_\mathrm{T} > 25$~GeV, and the event to have
$E_\mathrm{T}^\mathrm{miss} > 25$~GeV. 
Figure~\ref{fig12} shows the transverse mass distribution for the two
channels.

\begin{figure}[htb]
\begin{center}
\includegraphics[width=7.8cm]{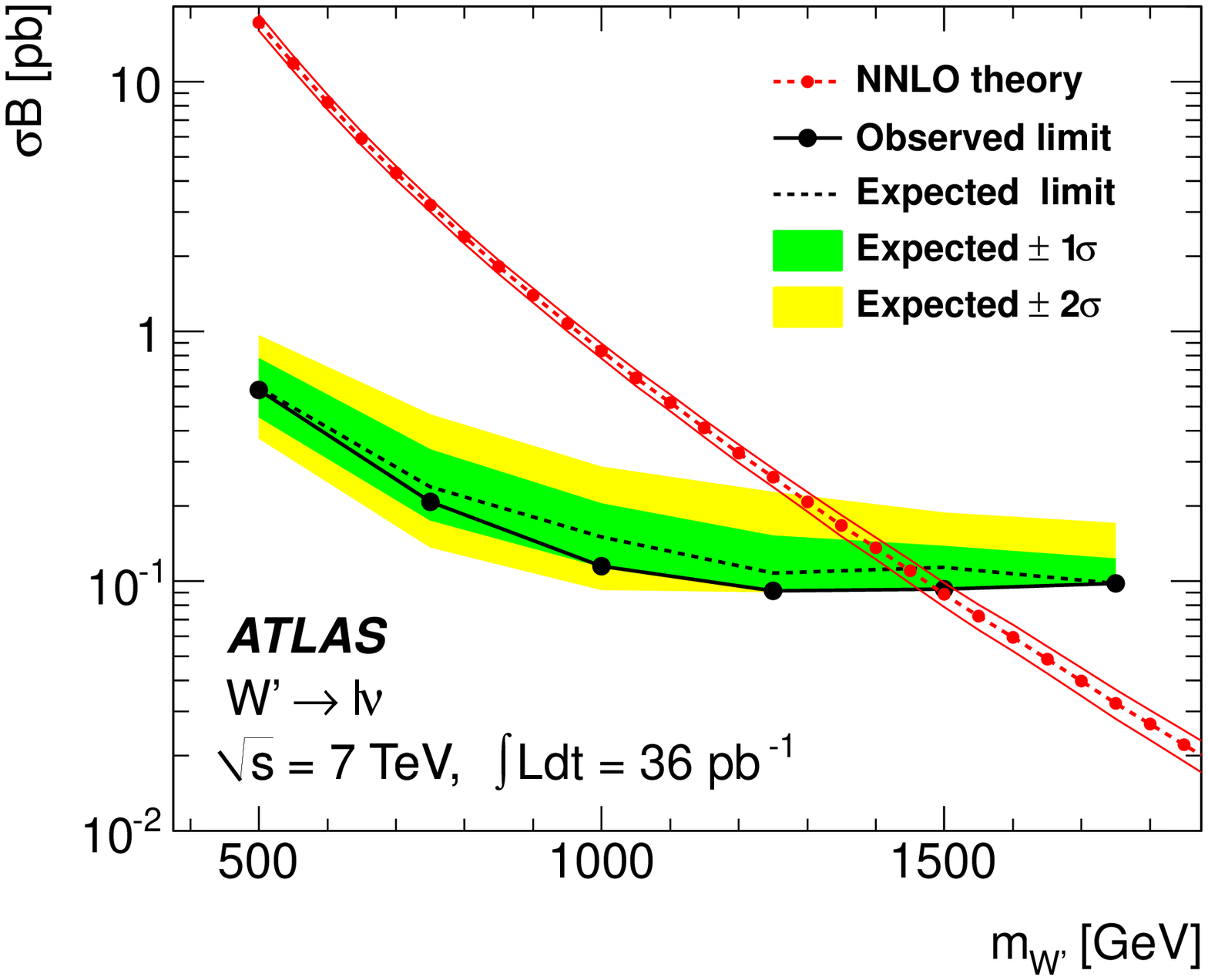}
\includegraphics[width=7.8cm]{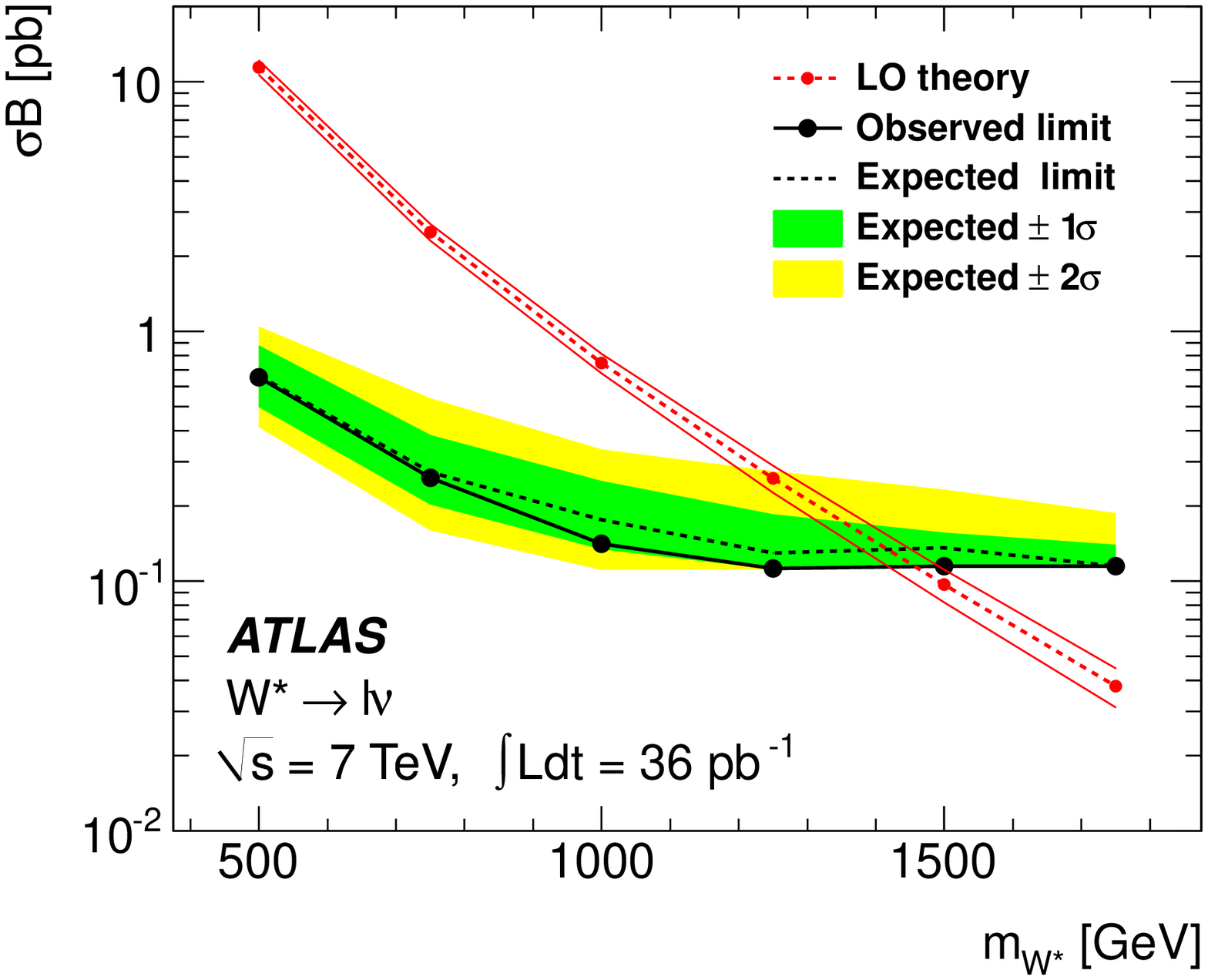}
\caption[]{Limits at the 95\% C.L. for W$^\prime$ (left) and W$^*$ (right)
production in the combination of decay channels.
The solid lines show the observed limits with all uncertainties.
The expected limit is indicated with dashed lines surrounded by
$1\sigma$ and $2\sigma$ shaded bands.
Dashed lines show the theory predictions (NNLO for W$^\prime$, LO for
W$^*$) between solid lines indicating their uncertainties.
\label{fig14}}
\end{center}
\end{figure}

The agreement between data and the expected backgrounds is good. 
Limits on $\sigma B$ for each W$^\prime$ and W$^*$ mass and decay
channel are set using a likelihood function as input to the estimate
$\mathrm{CL}_s = \mathrm{CL}_{s+b}/\mathrm{CL}_s$.
To set limits, we counted events with $m_\mathrm{T} > 0.5\;
m_\mathrm{W^\prime/W^*}$. 
Figure~\ref{fig14} shows the limits on the cross section times branching
ratio.
A W$^\prime$ with mass below 1.49~TeV and W$^*$ with mass below 1.47~TeV
are excluded.

\section{High Mass Dilepton Resonances}

\begin{figure}[htb]
\begin{center}
\includegraphics[width=7.8cm]{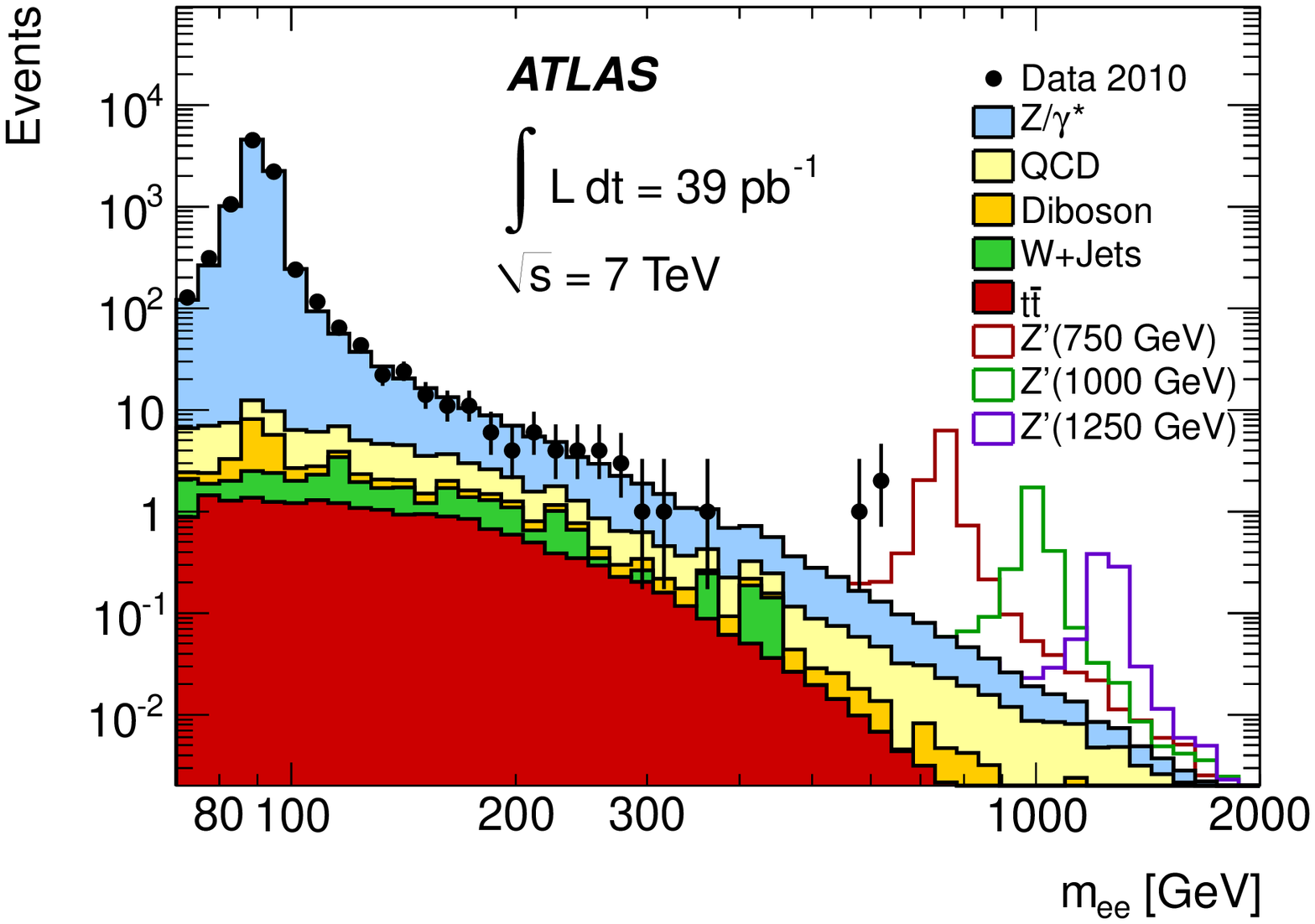}
\includegraphics[width=7.8cm]{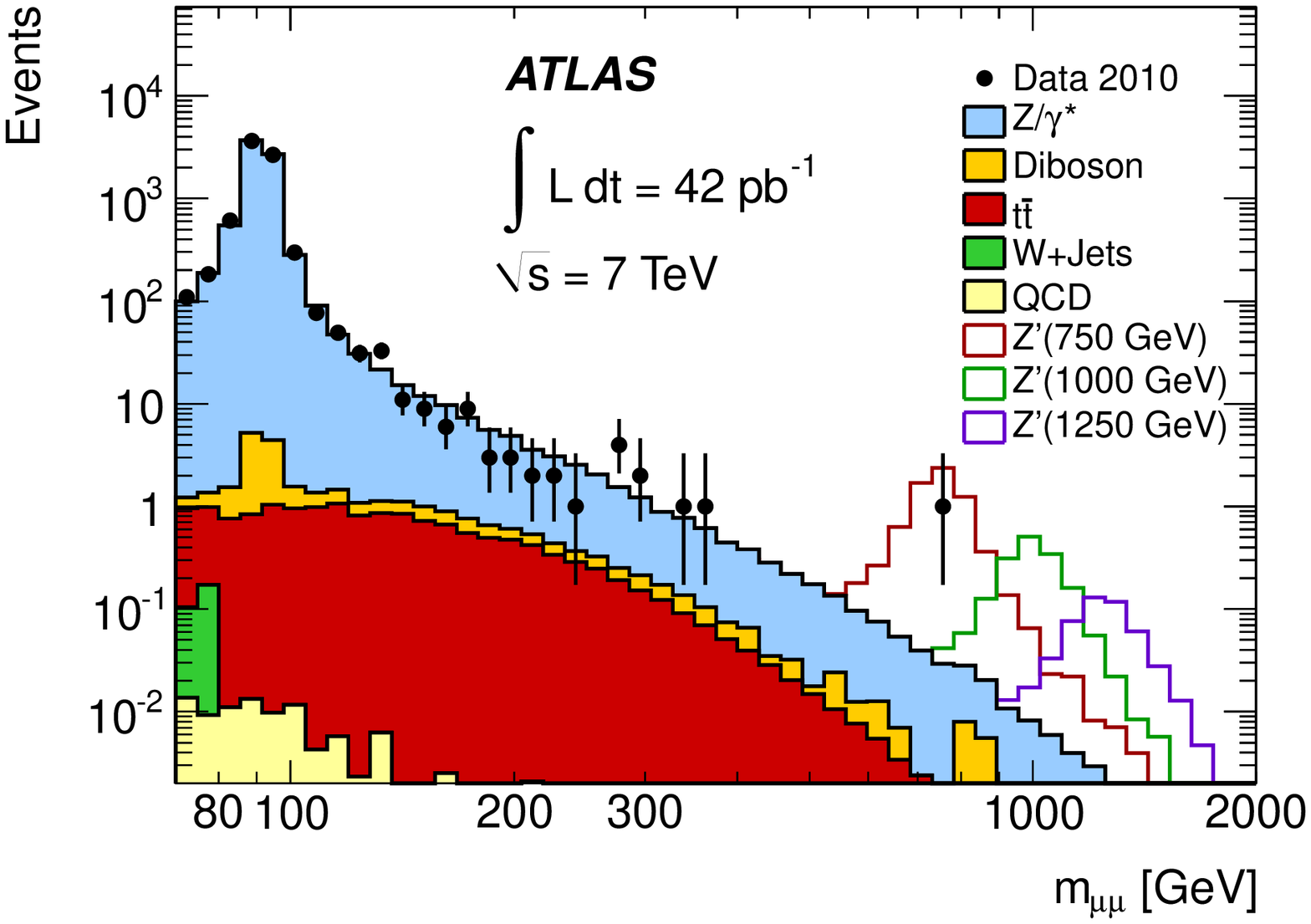}
\caption[]{Dielectron (left) and dimuon (right) invariant mass
distribution after final selection, compared to the stacked sum of all
expected backgrounds, with three example Z$^\prime_\mathrm{SSM}$ signals 
overlaid. 
\label{fig16}}
\end{center}
\end{figure}

ATLAS has performed a search for high-mass neutral resonance
states decaying to two leptons of the same generation.~\cite{Zprime} 
Examples of such high-mass resonances are 
new heavy neutral gauge bosons (Z$^\prime$ and Z$^*$),
the Randall-Sundrum spin-2 graviton, and
a spin-1 techni-meson.
The search looked for Z$^\prime \to \mathrm{e^+e^-}$ or $\mu^+\mu^-$,
where Z$^\prime$ is a high-mass sequential SM (SSM) gauge boson
with SM couplings, or a Z$^\prime$ motivated by an $E_6$ model.
Six different $E_6$ motivated gauge bosons were searched for with
different mixing angles between the two $U(1)$ states.
We assumed the resonance has a narrow intrinsic width compared to the
detector mass resolution, and required
$E_\mathrm{T} > 25$~GeV for the electrons in the dielectron channel, and
$p_\mathrm{T} > 25$~GeV for the muons in the dimuon channel.
Figure~\ref{fig16} shows the invariant mass distributions for the two
decay channels.

\begin{figure}[htb]
\begin{center}
\includegraphics[width=7.8cm]{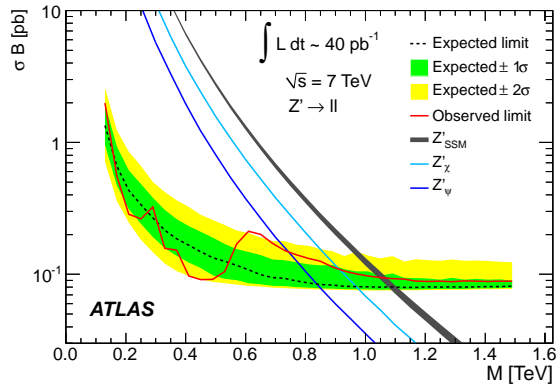}
\caption{Expected and observed 95\% C.L. limits on $\sigma B$ and
expected $\sigma B$ for Z$^\prime_\mathrm{SSM}$ production and two
$E_6$ motivated Z$^\prime$ models with lowest and highest $\sigma B$ for
the combined electron and muon channels.  
The thickness of the SSM curve represents the theoretical uncertainty.
\label{fig18}}
\end{center}
\end{figure}

Given the absence of a signal, an upper limit on the number of
Z$^\prime$ events is determined using a Bayesian approach.
For each Z$^\prime$ pole mass, a uniform prior in the Z$^\prime$ cross
section was used.
Figure~\ref{fig18} shows the limits on cross section times branching
ratio for the combined decay channels.
The measured and expected (shown in parenthesis) lower mass limits are 
0.957 (0.964)~TeV in the dielectron, 0.834 (0.895)~TeV in the dimuon,
and 1.048 (1.084)~TeV in the combined decay channels.
The lower mass limits on the $E_6$ gauge bosons range from 0.738~TeV
(Z$^\prime_\psi$) to 0.900~TeV (Z$^\prime_\chi$).

\end{document}